# Hydrological collapse in southern Spain under expanding irrigated agriculture: Meteorological, hydrological, and structural drought

Victoria Junquera*[1,2], Daniel I. Rubenstein[2], Simon A. Levin[2], José I. Hormaza[3], Iñaki Vadillo Pérez[4], and Pablo Jiménez Gavilán[4]

[1] High Meadows Environmental Institute, Princeton University, Princeton, NJ, USA

[2] Department of Ecology and Evolutionary Biology, Princeton University, Princeton, NJ, USA

[3] Institute for Mediterranean and Subtropical Horticulture La Mayora (IHSM La Mayora-UMA-CSIC), Algarrobo-Costa, Málaga, 29750, Spain

[4] Department of Geology, Faculty of Science, University of Malaga, 29071 Málaga, Spain

*Victoria Junquera

**Email:**  vjunquera@princeton.edu

**ORCID IDs:**

Victoria Junquera: 0000-0003-0402-3659
Daniel I. Rubenstein: 0000-0001-9049-5219
Simon A. Levin: 0000-0002-8216-5639
José I. Hormaza: 0000-0001-5449-7444
Iñaki Vadillo Pérez: 0000-0002-9767-6618
Pablo Jiménez Gavilán: 0000-0002-4042-0478

**Abstract**

Spain is the largest producer of avocado and mango fruits in Europe. The majority of production is concentrated in the Axarquía region in the south, where subtropical fruit plantations and associated water demands have steadily increased over the last two decades. Between 2019-2024, the region underwent an extreme water crisis. Reservoir reserves became nearly depleted and groundwater levels dropped to sea level in several locations, where seawater intrusion is likely, causing large socioeconomic impacts including short-term harvest losses and a long-term loss in economic centrality. We examine the causal pathway that led to this crisis using a mixed-methods approach, combining data from key informant interviews, an exhaustive review of legal documents, and quantitative analysis of time series and spatially explicit data. In particular, we analyze dam water use for irrigation and urban use, meteorological data, reservoir and groundwater levels, and irrigation land cover maps. Our findings show that an unusual meteorological drought was the immediate cause for the decline in reservoir and groundwater reserves (hydrological drought), but the underlying cause was a chronic and structural long-term imbalance between water demand and resources resulting from several structural governance shortcomings: large uncertainties in water resource availability and use hampering effective planning, lack of enforcement of individual water quotas, and the absence of regulatory mechanisms to flexibly impose resource use restrictions at both micro and macro levels based on the overall resources of the system. We propose concrete policy interventions aimed at sustainably enhancing the resilience of the system that can be useful to efficiently manage water shortages in other regions with similar problems.

**Significance Statement**

Our work provides important insights for the management of global freshwater-stressed areas. We apply a complex systems perspective and identify the water crisis in Axarquía as a hydrological collapse in the sense of a social-ecological regime shift. We show that emergency management measures were temporarily effective in preventing a complete depletion of reservoir resources and limiting socioeconomic damages, but these measures did not enhance the system's social-ecological resilience. Our analysis illustrates the interlinkages between social, institutional, and natural systems, showing that the crisis was caused by a confluence of exogenous and endogenous short- and long-term factors. Accordingly, we argue that sustainable solutions must consider all these scales and systems.

**Introduction**

Irrigation is the primary consumer of freshwater globally (1) and a major driver in global water storage variability and stress (2). Most of the currently stressed basins that are simultaneously losing freshwater storage are agriculturally significant and heavily irrigated (3). The Mediterranean basin, in particular, exhibits high levels of freshwater stress (3) and accelerated climate change and land transformations (4, 5). Addressing freshwater scarcity requires an integrated approach that includes not only traditional supply-side solutions, but also alternative supply options, demand management, and improved water management and economic systems (6). Conversely, a confluence of climatic factors and inadequate management can lead to extreme water crises. The Axarquía region in southern Spain is a case in point.

Over the last several decades, Spain has positioned itself as a major producer and exporter of subtropical fruits in Europe, producing 117,000 tons of avocado (*Persea americana* Mill.) and 41,000 tons of mango (*Mangifera indica* L.) in 2021 (7), constituting, respectively, 15% and 9% of net avocado and mango imports in Europe (8). Roughly 40% of Spanish avocado plantations (8000 ha) and 90% of mango plantations (4250 ha) are located in Málaga Province (7) concentrated in the Axarquía region. Subtropical fruit trees in Axarquía were first planted in irrigated areas around river basins, displacing mostly horticultural crops with comparable water requirements, but subsequently expanded into higher-elevation rainfed areas, significantly increasing the overall water demand (9, 10). A large water reservoir, the Viñuela dam, was built in the 1990s to ensure a continuous water supply for urban and agricultural uses. In October 2022, following several years of below-average precipitation, the dam's water levels dropped to near 10%. To avoid falling below the 8% threshold at which the dam becomes inoperable, authorities shut off the supply of irrigation water for the first time in the dam's history and imposed restrictions on urban water supply. Groundwater in the main coastal aquifer, the Vélez, reached the lowest recorded levels since monitoring began in the late 1990s and dropped to sea level in several locations.

In January 2024, dam water use restrictions were still in effect and dam levels reached a historical low of 7.5% (11, 12). The largest urban centers prospected for alternative ground- and surface water resources in neighboring regions (13). The large cuts in irrigation water supply and below-average rainfall had a major impact on fruit production in 2022/23 and resulted in harvest losses of 80% for mango and 50% for avocado compared to the previous year (14), causing important tree mortality in orchards where no water for irrigation was available. Local media labeled these developments a hydrological collapse (15), which had been anticipated by local environmental organizations since 2017 (9).

How could a prosperous agricultural region endowed with a modern water infrastructure decline into a state of hydrological collapse? To answer this question, we apply a multi-disciplinary perspective, jointly examining meteorological variables, water resources, use, and demand, and land and water governance. We use a mixed-methods approach combining qualitative data from key informant interviews with quantitative analysis of time series and spatial data, as well as a detailed review of legal documents, to identify causal mechanisms and pathways (16, 17) underlying the hydrological dynamics in Axarquía. As part of this process, we examine causal relations explicitly mentioned in the interviews, relationships between system variables, as well as temporal correlations between exogenous triggers and system variables.

In our analysis, we distinguish between three forms of drought. *Meteorological drought* (also known as climatological, rainfall, or pluviometric drought) refers to a period of below-average rainfall (18). *Hydrological drought* refers to a lack of water in the hydrological system, manifesting as abnormally low streamflow or levels of rivers, lakes, reservoirs, and groundwater (19). The term *structural drought* is generally not included in drought typologies despite its relevance in water stressed areas (20). It refers to chronic—hence, structural—water scarcity caused by a long-term imbalance between water demand and available water resources, resulting in a so-called structural water deficit (21, 22). Structural drought refers to a situation of recurrent hydrological droughts and it is generally attributable to mismanagement (23).

In the Results section, we examine climatological patterns over the last two decades to show how a meteorological drought turned into a hydrological drought when available water resources could not meet the growing water demand associated with expanding irrigated agriculture. In the Discussion, we highlight the main water and land governance shortcomings that underlie the structural water deficit in Axarquía and suggest concrete interventions.

**Study area**

**Geoclimatic conditions and land use history:** Axarquía is a county in southern Spain's Málaga Province within the Autonomous Community of Andalucía (Figure 1A). It extends across 1,026 km$^2$ (102,600 ha) and includes 31 municipalities with 220,500 registered residents in 2021 (24), a 44% increase since 2001 (25). Geographically, Axarquía is delimited by mountain ranges in the north (*Sierra de Tejeda y Almijara*) and west (*Montes de Málaga*) and the Mediterranean Sea in the south, with elevations ranging from 0 to 1200 meters above sea level (masl). Main rivers include the Vélez, Algarrobo, Torrox, Chíllar, and Arroyo de la Miel (Fig. 1C). Climate is subtropical Mediterranean with an average annual temperature of 18.9 ºC and average annual precipitation of 429 mm, concentrated during the wet season from September to May (both values measured at Viñuela reservoir; Section 3). Temperatures rarely drop below freezing at lower elevations, making the coastal areas suitable for subtropical fruit production. Agriculture and tourism are the two most important economic pillars of Axarquía, with agricultural activities comprising 11% of economic output in 2021 (26). The region has a long history of irrigation dating back to the Moorish period (9). In the 1960s, agricultural production was divided between irrigated crops such as tomatoes, potatoes, and the first avocado trees in the alluvial plains, and rainfed (non-irrigated) crops such as vine, citrus, almond, olive, or carob trees in the foothills (9). Land use became gradually less intensive and interspersed with saltus (scrubland) towards higher elevations, followed by natural vegetation in the mountain ranges (now largely protected areas) (9, 27). By 1988, there were 1,790 ha of avocado, up from around 300 ha a decade earlier, but adoption of mango remained insignificant (9). The construction of the Viñuela reservoir system in the 1990s provided a stimulus for the production of irrigated crops, in particular subtropical fruit plantations, which progressively expanded into higher-elevation and traditionally rainfed areas.

**Water resources:** The hydrological regime of Axarquía is determined by the Viñuela reservoir system, built as part of the 1989 Guaro Irrigation Plan and Water Supply to the Costa del Sol (28), also known as *Plan Guaro* (PG). The PG aimed to ensure a steady supply of water for agricultural and urban uses during prolonged droughts, which are frequent in the region (29). The Viñuela Reservoir System comprises the La Viñuela dam (164 hm3 capacity) built on the Guaro River (known downstream as the Vélez River) at 139 masl (at its base) and completed in 1989, as well as eight ancillary diversion dams on Guaro River tributaries, completed in 1995 (29, 30). The PG restricts the use of reservoir water to areas situated below 140 masl (28). This comprises 8,900 ha (4,870 ha of which were already irrigated in 1989) divided into nine sectors for management purposes (Fig. 1B). Water distribution pipelines were built in the late 1990s for irrigation Sectors 1-5, known as left-side sectors, and in 2014-2015 for right-side sectors 6-8 (31) (Fig. 1B). Groundwater bodies include detritic aquifers in the Vélez and Torrox basins with annual recharge rates around 40 mm; low-permeability metapelite (Metapelitas) and flysch (Colmenar-Periana) aquifers at mid-range elevations with low recharge rates (15 mm); and carbonate aquifers in mountain ranges with high recharge rates (250 mm) (9) (Fig. 1D). Surface water resources include rivers and springs. In November 2021, a project was completed to transfer 5.2 hm$^3$/year of tertiary treated wastewater (so-called regenerated water) from the Vélez-Málaga wastewater treatment to Sectors 6-8 (31), comprising 2,700 ha (32, 33). The regenerated wastewater capacity was subsequently increased to 22.5 hm$^3$/year by April 2024 by retrofitting four additional treatment plants (34, 35).

**Water governance:** The general principles of water management in Spain were established in the 1985 Water Act (36), which declared all surface waters and for the first time also all renewable groundwater bodies as public goods (37, 38) and created a Water Registry for permits and concessions (39, 40). The Water Act was updated in 2001-2003 to incorporate the requirements of the European Union Water Framework Directive (WFD) (41). The WFD commits EU member states to achieve good qualitative and quantitative status of all water bodies and requires the implementation of six-year water management plans (hydrological plans), including progress evaluations and reporting. The last three hydrological plans follow the same calculation methodology (42) in alignment with the WFD.

Water management and planning are the responsibility of River Basin Authorities (*Confederaciones hidrográficas*), which are administrative entities comprising one or more river basins. Units whose boundaries fall entirely within one Autonomous Community are managed at the regional rather than at the national level (30, 39). This is the case for the Andalusian Mediterranean River Basin District (*Demarcación Hidrográfica de las Cuencas Mediterráneas Andaluzas*, DHCMA), which falls under the purview of the regional government of Andalucía (*Junta de Andalucía,* JdA).The DHCMA is divided into 6 Systems and 16 Subsystems (30). System II, comprising Subsystems II-1, II-2, and II-3, corresponds to the Axarquía region.

Under the Water Act, private groundwater uses before 1986 could be transformed into a concession or, alternatively, remain in the private property domain by registering the use within three years in the Registry of Private Waters or within six years in the Catalog of Private Waters. Most users opted for grandfathering, as a result of which most groundwater use in Spain remains within the private domain and does not require a concession (43). New groundwater exploitations after 1 January 1986 are subject to an administrative permitting process. Extractions below 7,000 m$^3$/year merely require an inscription. All other groundwater uses and all non-groundwater uses must have a concession (44). Insufficient resources and unclear responsibilities caused many river basin authorities, including the DHCMA, to severely fall behind in processing permits (38, 45). Both the Catalog and the Registry remain incomplete to this date, particularly regarding private uses (39, 46). Almost forty years after the 1985 Water Act, possibly up to 80-90% of water wells in Spain are illegal or illicit (39, 43).

Many water users belong to local water management collectives known as Irrigation Communities (ICs), which have a long tradition and were originally set up around common irrigation infrastructures such as canals (47). ICs tend to have a narrow spatial scope and rarely comprise the majority of users within one aquifer, even though the WFD mandates this for overexploited aquifers (39). In Axarquía, there are 114 ICs, most of them private, with access to different water sources; some ICs rely almost exclusively on dam water, whereas others (e.g. outside the PG boundary), rely primarily on ground- or surface water.

**Land-use governance:** The Spanish Constitution of 1978 assigns responsibility for spatial planning to the regional governments (48). The main land-use plans are Municipal Master Plans, with a primary focus on designating suitability for urban development and protected status (48). A key deficiency of territorial planning in Spain is that it is not integrated with water resources planning (49) despite the fact that the WFD fosters the principles of Integrated Water Resources Management (IWRM) (50). Water resources and spatial planning are mostly regarded as two separate spheres (51), resulting in an institutional mismatch in the governance of land and water resources: whereas water planning has its administrative scope in the hydrological basin, land use planning falls under the responsibility of the region, subregions, and municipalities (51).

**Economics of irrigation:** The cost of irrigation water in Axarquía varies substantially depending on the water mix, depth to groundwater, and IC affiliation. The average cost of irrigation water in the DHCMA was 0.15 Euros/m$^3$ in 2006 (52). Viñuela reservoir water tariffs are around four cents per cubic meter (0.036 Euros/m$^3$) for irrigation and 11 cents per cubic meter for urban water in 2024 (53). However, individual users must be affiliated with an IC in order to gain access to reservoir water. IC membership entails additional costs and higher tariffs factoring in amortization and overhead. As an example, the IC Huertas Bajas charges 0.13 Euros/m$^3$ for the first 600 m$^3$, 0.2 Euros/m$^3$ for the next 150 m$^3$, and 0.3 Euros/m$^3$ for higher volumes. While the use of groundwater is free, it involves pumping costs, averaging 0.25 Euros/m$^3$ in the DHCMA (54), more than double the Spanish average (39, 52). Since 2022, users in Sectors 6-8 have access to regenerated wastewater at a cost of roughly 0.40 Euros/m$^3$.

The yield of avocado plantations in Málaga Province in 2017-2021 was 7.4±0.6 tons/ha, and 6.8±1.3 tons/ha for mango (55). However, this includes plantations of all ages, including those that are not yet productive. (56) report an average yield of 11 tons/ha in a commercial mature avocado plantation in Axarquía under conventional irrigation practices (6500 m3/ha) corresponding to 82% of theoretical water needs. (57) report an average yield of 14.5 tons/ha in an experimental mango farm near Axarquía (under similar geoclimatic conditions) irrigated at 100% of ETc corresponding to 4,750 m$^3$/ha. In practice, the regulatory water allocation for irrigating avocado and mango plantations is 5,300 m$^3$/year (54) and typical water use in mango plantations is 4500-5000 m$^3$/year. Considering a unit price of 0.25 Euros/m$^3$, the cost of irrigation is around 1,300 Euros/ha for avocado and slightly lower for mango.

**Results**

**Drought events:** SPI-12 values show several multi-year droughts (Fig. 2). Given that our precipitation data start in October 1995, they only capture the end of the important 1992-1996 drought. Drought severity values show that major droughts took place in 2005-2009, 2014-2018, and 2019-2023, which extended into 2024 (58). The shorter-term SPI-6 index shows that rain events during drought periods had lower frequency and lower precipitation values than during non-drought periods (Fig. S2-2). Median temperatures exhibit a slight upward trend between 2013-2022 (Fig. S2-1).

**Extent of irrigated agriculture:** The area of avocado and mango plantations in Málaga Province exhibits discrepancies between government (59) and private sector (55) estimates (Fig. 3). Averaged together, in 2021 there were 7,000 ha of avocado and 4,100 ha of mango in Málaga, up from 4,500 ha and 500 ha, respectively, in 1999. The irrigated surface area in Axarquía in 2018 (Fig. 4A) comprised 16,000 ha (+11% since 2009), 8,660 ha of which were avocado and mango plantations (+18% since 2009). Taken together, these data suggest that around 85% of avocado and mango plantations in Málaga Province are located in Axarquía. (9) developed a LC map of Axarquía in 2017 based on supervised classification of remote sensing data, identifying 6,400 ha of avocado and 3,500 ha of mango in Subsystem II. These values combined exceed Tragsatec's estimate by 1,000 ha and come close to the total values for the whole province, indicating that they might be overestimated (Fig. 3). Avocado and mango plantations expanded rapidly in the early 2000s (Fig. 3), following the extension of dam irrigation infrastructure to right-side sectors (Fig. 1B) in 1999. Between 2005-2014, mango plantations spread faster despite sharp fluctuations in mango market prices (Fig. 3). The expansion of both crops slowed down during the droughts of 2014-2018 and 2019-2023 and surged in between (Fig. 3). The comparatively faster expansion of mango in the last two decades is attributable to its lower water requirements and higher drought resistance. Mango is preferred at higher elevations, especially in areas outside the PG boundary where use of dam water for irrigation is theoretically not allowed and aquifers (notably the Metapelitas) have low yields. In practice, dam water use has extended beyond the PG regulatory perimeter (Fig. 4B), as individual users have illicitly extended pipelines from the dam water distribution network into higher elevations, "tens or even hundreds" of meters above 140 masl. Already in 2008, 2,840 ha outside the PG boundary (and 3,660 ha within) were irrigated with dam water (60).

**Water balance in hydrological plans:** The hydrological plans include estimates of actual water use by source (dam, surface, ground-, and regenerated water) and theoretical demand by sector (irrigation, urban, industrial, and golf). The balance between water use and demand in Subsystems II-1&3 is negative in all four hydrological plans, indicating that demand is higher than use (Table 1). Whereas the first three plans attribute this difference to "under-allocation of irrigation water" and "demand not met" (60) (implying that farmers used less water than needed and crop growth was thus suboptimal), the most recent plan acknowledges that the difference is attributable to overexploitation of groundwater resources (61), meaning that actual groundwater use is likely 11 hm$^3$ higher than the official estimate. Despite this acknowledgement, the plan does not correct the presumably inaccurate estimates.

Other aspects also stand out in Table 1. Surface water use estimates differ substantially between planning cycles. The main reason for this is that surface water use is calculated as the balance of total water *demand* minus all other water *uses* (dam, groundwater, and regenerated). This points to i) a circular reference in the calculations and ii) calculation of water use based on theoretical demand rather than direct estimation. Moreover, groundwater use was significantly revised upward in the last planning cycle, from 31 to 45 hm$^3$, which we are able to replicate (S3). Such a large difference suggests that groundwater use was underestimated in all previous planning cycles.

**A closer look at groundwater balance:** Hydrological plans include aquifer-specific information on groundwater use, permitting volume and status, and exploitation index (the ratio of total extraction over sustainable extraction rate), summarized in Table 2 for the three aquifers in SII-1&3. Several elements stand out in this table. Groundwater use from the Vélez aquifer dropped substantially between PH 1998 and PH 2009-2015, associated with a decrease in irrigated area from 4,505 to 2,578 ha (Table S2-1). This decline reflects the start of operation of the dam in 1998-1999, allowing users to use dam water instead of relying solely on groundwater, since dam water is preferred for its superior hydrochemical quality compared to Vélez aquifer water. Total groundwater use increased markedly in the last planning cycle, from 17 $hm^3$ to 33 $hm^3$ (Table 2). This reflects lower availability and use of dam water compared to the previous cycle (-5 $hm^3$) and an upward revision of irrigated area and irrigation intensity (Table 1). Accordingly, the aquifer exploitation index increased for all three aquifers and for the first time exceeded 1.0 for the Vélez and Metapelitas aquifers, indicating unsustainable exploitation.

The maximum sustainable extraction rate (a value specific to each aquifer based on aquifer recharge rates and rainfall), used to calculate the exploitation index, nearly doubled for the Torrox and Metapelitas aquifers in the last planning cycle based on an upward estimation of infiltration rates (Table 2, RS). This is surprising because the calculation methodology and aquifer characteristics did not change during that period.

The volume of registered groundwater permits increased substantially in the last plan for all three aquifers —for example in the Vélez from 8.6 $hm^3$ to 14 $hm^3$. This is presumably due to a reduction in the permitting backlogs during that time, as pressure on the regional government intensified because of the difficult hydrological situation. For the Metapelitas aquifer, groundwater permits have been processed and approved for 19 $hm^3$, with additional 2.8 $hm^3$ under review. Yet estimated groundwater use -the value that is used to calculate the exploitation index- is only 9.5 $hm^3$. While permitted volumes represent maximal allocations and can therefore exceed actual use rates, it is questionable that less than half of the permitted volume is used in a context of water scarcity.

Conversely, groundwater use from the Vélez aquifer (23 $hm^3$) is higher than the permitted volume, including permits under review (20 $hm^3$). The difference suggests that Vélez aquifer water is used beyond its geological perimeter (e.g. by extending pipelines), which is allowed. To elucidate this, we compare the extent of irrigated area located *above the hydrological boundary* of each aquifer in 2009 (our estimate), with the extent of irrigated area *using water extracted* from each aquifer (reported in PH 2009-2015) (Table S2-1). 1,325 ha were irrigated above the Vélez aquifer boundary, but 2,578 ha were reportedly irrigated with Vélez groundwater, implying that Vélez groundwater was used in 1,200 ha outside its perimeter. The situation is reversed for the Metapelitas aquifer: 4,877 ha were irrigated above its perimeter, but only 764 ha were reportedly irrigated with its water (Table S2-1), suggesting that over 4,000 ha situated above the Metapelitas aquifer were irrigated with other sources. Presumably, 1,200 ha out of these 4,000 ha were irrigated with Vélez aquifer water. The rest (2,800 ha) must obtain water from other sources, notably surface water and, as Fig. 4B shows, dam water. In fact, 2,000 ha located above the Metapelitas aquifer fall within the PG boundary and are entitled to use dam water. Assuming an average irrigation rate of 5,000 $m^3$/ha for subtropical plantations, the remaining 800 ha would require approximately 4 hm3, corresponding to 42% of the official groundwater use estimate of 9.5 $hm^3$. This 4 $hm^3$ could come from i) dam water illicitly pumped outside the PG boundary and/or ii) water extracted from the Metapelitas aquifer but not reported. Considering that the official groundwater use estimate already exceeds the sustainable extraction rate (Table 2), taken together these data suggest that the Metapelitas aquifer is overexploited, and that significant water volumes are not accounted for in the water balance.

**Reservoir levels**: Viñuela dam water levels fluctuate within and between years, with levels below 30% indicating a situation of water stress, and below 15% denoting a state of emergency (62). The level dropped below 10% in 1996, 2009, and 2023 (Fig. 5). The juxtaposition of precipitation data and dam parameters (Fig. 6) shows that the immediate cause for the two most recent drops was a reduction in inflows to the dam (Fig. 6A) owing to two meteorological drought events in close succession (Fig. 2). Inflows in 2018 were not sufficient to restore the volumes abstracted from the dam during prior and subsequent droughts (Fig. 5). Despite steadily declining dam water levels between 2013–2017, dam water use for irrigation was above average and increased during that period. Overall use was reduced in 2016 and 2017 mostly by decreasing urban water supply (Fig. 6B). From 2019 to 2021, dam water use increased again (Fig. 6B) while dam levels declined from 47% to 28% (Fig. 4). Despite the severity of the situation, dam water use in 2021 exceeded that of the previous year (Fig. 6B). Between the years 2013–2022 (except 2018) the dam operated at a negative balance, with water use exceeding net recharge (Fig. 6C), which explains the progressive water level decline, as well as the growing divergence between resource availability and use (Fig. S2-3).

**Groundwater levels:** During meteorological droughts, groundwater use typically increases to compensate for lower reservoir levels. Lower precipitation also translates into slower aquifer recharge. These two factors combined —higher extraction rate and slower recharge rate—explain the drop in groundwater levels during meteorological droughts (43). Between 1997 and 2013, Vélez aquifer levels oscillated between 90% and 60% (as measured by a representative piezometer located in the center of the aquifer; see S2-4 for all piezometer readings). The lowest level was registered during the severe 2005-2009 drought. Nevertheless, aquifer exploitation rates were relatively sustainable until around 2016. In 2014, Vélez groundwater levels entered a period of steady decline, which was intensified after 2019. By 2023, 9 of the 11 piezometers located in the coastal sector of the aquifer (Fig. S2-4) had reached levels close to 0 masl, at which saline seawater intrusion may become a problem.

To understand the dynamics of the Vélez aquifer in recent years, we compare the last two major droughts. From January 2014 to December 2017, aquifer levels declined from 82% to 46%, with cumulative rainfall averaging 326 mm/year, and dam water use averaging 37 $hm^3$/yr, out of which 22 $hm^3$/year were for irrigation. From January 2019 to December 2021, aquifer levels dropped from 81% to 15%, with cumulative rainfall averaging 300 mm/yr, and dam water use averaging 34 $hm^3$/year (20 $hm^3$/year for irrigation). While accurate water cycle assessment requires daily resolution and evapotranspiration consideration, the numbers above illustrate that the period of groundwater level decline after 2019 coincides with below-average precipitation and dam water use, both of which would have increased groundwater demand. Nevertheless, the rate of decline in groundwater levels is higher than the rate of decline in dam water use or rainfall after 2019 (Fig. S2-3), suggesting that other factors were also involved.

**Discussion**

We identify the main drivers and causal pathway underlying the 2019-2024 hydrological crisis in the Axarquía social-ecological system based on an analysis of the principal time-varying variables influencing the hydrological cycle, detailed examination of water accounting and governance, and interviews with diverse stakeholders, and posit that this crisis constituted a collapse in the sense of a social-ecological regime shift. We pinpoint major sources of uncertainty that hamper effective management efforts and systemic governance shortcomings that thwart long-term resilience. We then propose policy interventions involving different leverage points within the system.

**Meteorological, hydrological, and structural droughts**: A main cause for the drastic drop of groundwater and dam levels in Axarquía was the occurrence of two major *meteorological* *droughts* in close succession, which constitutes a meteorological anomaly in this region. Because rain events were not absent but rather less intense and more spaced in time than during non-drought periods, drought severity becomes evident when assessed across intervals of 12 months and to a lesser extent 6 months, but is less apparent in shorter timeframes. This climatic pattern created an illusion of temporary relief during each rain event that could have delayed effective and determined management interventions. This sentiment was echoed by a local stakeholder who stated "I hope it does not rain" in the midst of the drought in 2021, because each rain event appeared to solve the chronic problem temporarily (63). The slow and reactive (64) initial management response can be attributed to socio-institutional barriers (65), such as the general disbelief that the drought would continue beyond 2020, let alone 2021, reluctance to intervene given the economic centrality of agriculture and tourism, and unclear emergency provisions.

In May 2021, the regional government declared a situation of "extraordinary hydrological drought" requiring the implementation of a drought management plan and transferring water management and planning responsibilities in the DHCMA to a special Drought Management Committee (*Comité de Gestión de la Sequía*) (66), which is authorized to enforce water conservation measures and make exceptional budget allocations (67). The measures imposed by the drought committee were significant and included reducing the dam water quota for irrigation from 5,160 $m^3$/ha to 2,000 $m^3$/ha in November 2021 (68) and to 1,500 $m^3$/ha in March 2022, in conjunction with a 20% reduction in urban water supply (69). As dam levels continued to drop and approached 10%, the irrigation supply was entirely shut off in November 2022 (69). This measure has remained in place until May 2024, when an emergency irrigation quota was approved for only 3 $hm^3$, corresponding to 500 $m^3$/ha (58) or roughly 10% of the long-term average.

The exploitation of groundwater resources intensified after 2019, and especially after 2021, as a direct result of reduced dam water supply and lower rainfall, both of which increase groundwater use. The continuous expansion of irrigated avocado and mango plantations, including the jump around 2018, presumably intensified the water demand. Moreover, the sudden and large intervention after 2021 may have increased groundwater pumping above and beyond a mere substitution of reduced dam water and rainfall, as water users strove to become more self-reliant by building new wells (67), which proliferated during this period. This suggests the existence of socio-institutional self-reinforcing feedback loops in groundwater extraction.

The combination of groundwater and reservoir resources is central to drought resilience (2). The steep drop in groundwater resources in Axarquía, evidenced by significant piezometric declines in many areas of the Vélez aquifer from 2019 onwards, signals the breakdown of the desirable dynamic where groundwater can compensate for lower surface water availability, reservoir reserves, and rainfall (43). The period after 2019 reflects the onset of a *hydrological drought* (19), with both reservoir and groundwater levels steadily declining and reaching critically low levels in 2023. The combination of dam and groundwater resources was insufficient to meet the water demand of the system, even after regenerated wastewater was introduced in October 2021 (54, 70). A clear indicator of the hydrological drought is the large harvest loss of avocado and mango fruit after 2020. This hydrological drought occurred within a context of *structural and chronic* water deficit, caused by an excessive—and steadily growing—water demand compared to available water resources in the region. Underlying this structural imbalance are major shortcomings in land and water governance (below).

**A collapse?** The dynamics after 2019 exhibit the characteristics of a social-ecological regime shift (71), involving a steady loss of resilience (72) caused by growing water demand, exogenous meteorological triggers (73), multi-scale feedbacks between the social and ecological systems (74), path-dependencies (75) such as the fractionation of water management among hundreds of irrigation communities, as well as a destabilizing effect of uncertainty and abrupt management changes (76). The system underwent nonlinear and rapid change, nearly depleting reservoir and groundwater resources within a few years. Some of the resulting social-ecological impacts are difficult to reverse (77)—most importantly, the region's reputational damage and loss in significance as a central locus of subtropical fruit production and trade in Europe, as avocado cultivation expands to other regions (such as western Andalucia, the Valencian Community in the East and the Algarve region in Portugal) and local trading companies increasingly rely on imports to meet growing demand and supply shortages (78).

**Governance shortcomings:** *(I) Uncertainty in water resource estimates:* The discrepancy between actual water resources and their quantification in hydrological planning is acknowledged in official documents (66, 79). This stems in part from a lack of scientific understanding about hydrogeological dynamics (44). The water balance calculations in the hydrological plans likely overestimate inflows to the dam, surface water flow, and aquifer recharge for two reasons. First, the calculations do not account for changes in land use and precipitation patterns over the last two decades.

The expansion of irrigated areas upstream and downstream of the dam basin increases evapotranspiration as well as extraction of surface and groundwater, all of which reduce dam and aquifer recharge rates (80–82). Second, changing rainfall patterns have caused a decline in so-called useful rain, which is the fraction of precipitation that effectively recharges surface and groundwater bodies. As rain events become more spaced in time and less intense, a larger fraction of the precipitation is "lost" to soil saturation (82). Despite this, aquifer recharge rates were surprisingly revised upward by 40% and 90%, respectively, for the Metapelitas and Torrox aquifers in the last hydrological plan.

Another major source of uncertainty stems from the lack of quantitative and qualitative monitoring in the Metapelitas and Torrox aquifers despite the fact that such monitoring is required under the WFD (EU, 2000 & 2006). Our results show sharp declining groundwater levels in the Vélez aquifer after 2019; arguably, the Torrox and Metapelitas aquifers underwent similar dynamics, but the absence of monitoring (a network of piezometers) makes verification impossible. Unlike the detritic Vélez and Torrox aquifers, the Metapelitas was long considered to have insufficient yield to support groundwater-irrigated agricultural development, leading local authorities to conclude that monitoring was not necessary. This has not changed despite the strong expansion of subtropical fruit plantations above the Metapelitas' perimeter over the last two decades.

*II. Uncertainty in water balance:* The attribution of a negative balance of 11 $hm^3$ to unaccounted groundwater overexploitation in the latest hydrological plan constitutes a clear sign of inaccuracies in water use and demand estimates. One reason is inadequate monitoring. Whereas *total dam* water flow rate is *continuously monitored and recorded* at the main distribution pipelines using high-accuracy flow meters, current regulations do not require reporting *individual* water use, let alone continuous recording. The vast majority of surface and groundwater permits merely require episodic proof (e.g., a water meter reading) that annual consumption is below the permitted threshold. Moreover, such proof is only requested if the user is subjected to a random inspection, which occurs rarely. Because extraction of surface and groundwater—jointly comprising 64% of total water use—is not reported, the managing authority must estimate these values based on irrigation community surveys of unknown accuracy and best-available information on spatially-explicit irrigated crop area and water use intensity (not directly measured).The significant upward revision of surface (+63%) and groundwater (+45%) use estimates may be attributable to third-party land cover maps based on remote sensing (83), but remote sensing is not routinely employed and such maps only exist for 2018 and 2009.

The discrepancies between permitted volume and use estimates, and between use and demand estimates, point at a potentially significant amount of irregular extraction, whether illegal (conducted without a permit or in violation of permit conditions) or illicit (e.g., if the permit request has been submitted but not yet approved). Our own assessment for the Metapelitas aquifer suggests that extraction could be around 40% higher (high-uncertainty estimate) *during normal meteorological periods*. During the last drought, the restriction and eventual shutoff of dam water supply deprived the system of roughly 30 $hm^3$, constituting around 30% of total water use, in a context of below-average rainfall. It is highly likely that this shortfall was at least in part compensated with groundwater. A recent investigation found 250 illegal wells in Axarquía (84, 85) and allegedly fraudulent groundwater extraction of up to 25 hm$^3$ (86), constituting 55% of official groundwater use estimates. The investigation was conducted by the Spanish environmental police (*Servicio de Protección de la Naturaleza*, SEPRONA), whose focus is on major environmental violations rather than routine inspections. Axarquía has only a handful of water compliance inspectors. The limited resource allocation for regulatory enforcement is a major cause for the pervasive unauthorized use of water (52).

**III. Lack of integrated land and water resources management:** The lack of integrated management means that *there is no regulatory mechanism to impose an upper bound on irrigation area* based on available water resources within the system. The Territorial Plan for Axarquía (79) makes an attempt at integrated management by calling for an adjustment of the Plan Guaro to the present-state reality of irrigated area sprawl (87), but these suggestions are not binding and have not been implemented (88). The sole administrative planning tool for irrigation, the National Irrigation Plan, does not dictate which areas may or may not be irrigated (89). As a result, owners of agricultural land can transform land from rainfed to irrigated as long as they have a water permit. In practice, many landowners submit a permit application (e.g., for a new well) and switch to an irrigated crop before the permit is processed, which can take years (52). Other possible administrative hurdles for switching to irrigation concern protected areas and land with natural woody vegetation, for which a deforestation permit is needed, but such cases only affect higher-elevation areas. However, there is no regulatory instrument to limit the extent of irrigated areas *at the scale of hydrogeological and water management dynamics*, in this case Subsystems II-1&3. This has led to an unsustainable expansion of irrigated crops in Axarquía and in many other regions in Spain (90).

*IV. The cost of water is (too) low:* The cost of irrigating avocado and mango plantations is around 1,300 Euros/ha. Based on yields reported in the literature in mature avocado and mango plantations (see above), and 2017-2021 market prices, namely 2.5±0.1 Euros/kg for avocado and 1.1±0.2 Euros/kg for mango (55), the estimated average gross income is around 27,500 Euros/ha for avocado and 16,000 Euros/ha for mango. Annual production costs are around 5,200 Euros/ha for avocado (91) and 6,500 Euros/ha for mango (92). The ratio of irrigation costs over net profit is thus around 13% for mango and 6% for avocado, compared to the 10% benchmark cited in studies of irrigation economics in Spain (39, 52). These figures illustrate why the cost of water has not traditionally factored into the decision of whether to establish an avocado plantation in Axarquía. Considering that rainfed crops such as olives or almonds were barely profitable in this region (9), the economic attractiveness of replacing them with either avocado or mango trees becomes apparent.

**Policy recommendations for sustainable water use:** Water resource management in Spain has traditionally focused on supply management. This is still the focus of short- and medium-term solutions recently proposed or implemented in Axarquía, namely increasing regenerated wastewater capacity, installing desalination plants (31), and building water transfer pipelines connecting Axarquía with other exploitation systems within the DHCMA, including a "water highway" to double water transport capacity from the western Costa del Sol (54, 70). Such measures can alleviate hydrological droughts in the short and medium term, but they are unlikely to resolve the chronic water scarcity and might worsen the situation in the long term by incentivizing further agricultural expansion (93). A sustainable course of action must address the structural water deficit and major governance shortcomings.

Policy interventions should aim to improve knowledge about available water resources and real water use in the region, and incorporate this knowledge into models and water balance calculations that inform hydrological management plans. As part of this, it is essential to implement a monitoring system for i) groundwater levels in each aquifer, ii) land cover in each parcel, and iii) water use at each extraction point. Mandatory monitoring and reporting by extraction point would entail additional (29) costs for individual users with a possible deterrent effect. Long-term, durable solutions will require a system-level approach to reduce the main cause for the structural water deficit, namely an excessive demand owing mostly to a continuous expansion of irrigated crops and to a lesser extent growing urban water demand (10). Land and water resources management should be integrated and adaptive (94) to account for the variable hydrological carrying capacity of the system and its changing water demands. The scale of management should be realigned to match hydrogeological dynamics. In practical terms, this means consolidating water management across resources (dam, ground, and surface) and users, and implementing flexible extraction quotas based on monitored resource stocks. Achieving this will require addressing institutional legacies that hamper adaptation, such as fractionated management across hundreds of irrigation communities, or rigid groundwater permit quotas, many of which are very high at 7,000 $m^3$/ha. Water demand could also be steered through water pricing. Finally, graduated sanctions and effective enforcement are essential to ensure regulatory compliance (95) and will require an adequate number of field inspectors, expedited permit processing, and strong commitment from the regional government. The additional costs associated with all of these measures, from the micro (individual) to the macro (system)-level, represent an internalization of presently external environmental and social costs of water use in the region (96).

**Limitations and future steps:** Certain causal attributions could be more robust through the additional use of models. Hydrogeological cycle modeling at one-day resolution accounting for temperature and evapotranspiration would allow to disentangle the relative effect of meteorological variables vs. water use on dam and aquifer levels, which would provide additional relevant information for management purposes.

**Conclusion:** Most social-ecological regime shifts are caused by a confluence of factors rather than by one single cause. Our analysis provides evidence of the chain of events that culminated in a state of hydrological collapse in Axarquía. An unusually long meteorological drought and slow management action caused a sustained drop in reservoir and groundwater reserves. Emergency measures effectively prevented the depletion of the dam to ensure the (limited) provision of urban water, but the abruptness and scope of the intervention may have exacerbated groundwater abstraction. The near-depletion of reservoir and groundwater resources (hydrological drought) within a short timespan was the result of exogenous meteorological triggers coupled with endogenous social-institutional shortcomings in land and water management that resulted in a growing structural imbalance between demand and resources—and loss of resilience—over decades. Durable and sustainable solutions require an integrated and flexible management approach that combines macro- and micro-level quotas, monitoring, and enforcement.

Axarquía is small, but it serves as an example for many agricultural regions in the Mediterranean and other areas that are characterized by increases in irrigated area coupled with a growing urban water demand. Similar challenges are faced globally (3, 97), for example in California (98, 99), the Murray-Darling Basin in Australia (100), Iran and other countries in the Middle East (2, 101), or Chilean fruit producing regions (102) where the combination of climatic variability, high water demand, and governance issues has led to significant water stress. Addressing these issues requires a holistic and integrated approach that balances the needs of agriculture, urban areas, and the environment to ensure long-term sustainability and resilience.

**Materials and Methods**

**Data collection:** We compiled monthly time series (see S1) of cumulative rainfall (mm) (January 1995-April 2023), average temperature at Viñuela (deg-C) (March 2012 – May 2023), Vélez aquifer level (masl) (March 1995 – May 2023), and Viñuela dam volume ($hm^3$) (January 1992 – April 2023). Other Viñuela dam parameters are only available for January 1992 until October 2022, including inflow (i.e., recharge), evaporation, and outflow (irrigation water supply, urban water supply, and discharged water) (all in hm$^3$).

We obtained annual data on avocado and mango production surface (ha) in Málaga Province from agricultural census data (1991–2021) (59) and from the Málaga Farmers' Association (2002–2021)(55), and annual average avocado and mango market prices (2002–2021) (55).

We obtained raster maps of irrigated land cover (LC) in Subsystems II-1&3 (10x10m resolution) for 2009 and 2018 (103, 104) developed by a consulting firm as part of a technical support study for the DHCMA (83). The study used automatic classification to assign LC categories to irrigated parcels based on satellite imagery (Landsat5 for 2009 and Sentinel2 for 2018) (83). The LC category Non-Citrus Fruit Trees comprises mostly avocados and mangos. For the years 2009 and 2018, we also obtain shapefiles of irrigated parcels i) included in the Irrigation Community Register (Elenco) (105), and ii) not in the IC Register (106). Additional spatial data include: perimeter of administrative regions (107); perimeters of Subsystems II-1, II-2 and II-3, groundwater bodies, and Plan Guaro (108); and a Digital Elevation Model (DEM) of Andalucía (109).

We examined all four Hydrological Plans (PH) that exist for the DHCMA. These are publicly available and include: PH 1998 (approved 1998) (110), PH 2009-2015 (approved 2011) (68), PH 2015-2021 (approved 2016) (111), and PH 2022-2027 (approved 2023) (54). Note that in PH 2021-2027, SII-3 is integrated within SII-1. PH 2015-2021 was repealed in May 2019 (112), thus PH 2009-2015 was in effect until 2023. We also obtained a spreadsheet containing detailed water accounting in PH 2009-2015.

Between December 2021 and October 2022, we conducted interviews with five producers; two members of an irrigation community with management and technical responsibilities; one member of the dam management committee; the owner of a local water well drilling company; representatives of ASAJA-Málaga and ASAJA-Valencia; the president of a local environmental organization; three high-level employees of different offices within the regional Department of Agriculture, Fishing, Water, and Rural Development (Consejería de Agricultura, Pesca, Agua y Desarrollo Rural de la Junta de Andalucía), including the Directorate of Planning and Water Resources (Direccion General de Planificación y Recursos Hídricos), and two persons at the Málaga Regional Office of Agriculture (Delegación de Agricultura); the president and a high-level manager, respectively, of two of the three main agricultural trading companies in Axarquía; a journalist of a local newspaper specializing on agriculture and irrigation; and three university professors from Universidad de Málaga focusing on irrigation and/or subtropical fruit production. Some persons were interviewed more than once. Overall, we conducted 24 interviews. Questions were open-ended and focused on past, present, and future dynamics and challenges.

**Data analysis**: To identify drought events, we calculated the Standardized Precipitation Index (SPI), a widely used drought index recommended by the World Meteorological Organization first introduced by McKee et al. (113). The SPI fits long-term precipitation values to a Gamma distribution and transforms them to a normal distribution (114). SPI can be calculated for different time scales. Shorter periods (1-3 months) are appropriate for assessing impacts on soil moisture, snowpack, or small river stream flows, and longer time scales best represent stream flow, reservoir storage (3-12 months) and groundwater recharge (12-24 months) (115). For arid climates where zero precipitation values are common, the SPI at short time scales (generally ≤3 months) is non-normally distributed and fails to indicate a drought occurrence (116). A drought event is defined when the SPI is continuously negative for 3 or more periods and drops below a certain threshold at least once; the event starts when SPI becomes negative and ends when it becomes positive, with the length of this period constituting the event duration (117). The lowest SPI value during the event is used to classify its drought class ($-1.0 < $ SPI $ \leq 0$ Mild; $-1.5 < $ SPI $ \leq -1.0$ Moderate; $-2.0 < $ SPI $ < -1.5$ Severe; $-2.0 \geq $ SPI Extreme). The severity of a drought event is defined as the cumulative SPI below a threshold (e.g., 0, −0.5, or −1) (117). We used monthly cumulative precipitation values to calculate 6- and 12-month SPI using the built in function spi from the package SPEI (118) in R. To calculate drought severity values we used the function RunDS from the package drought (119) with a threshold of 0.

To examine temperature trends, we use the nonparametric Mann-Kendall test in combination with the Theil-Sen median slope estimator (120), using the functions MannKendall (package Kendall) (121) and TheilSen (package robslopes) (122) in R.

We extracted water balance data (water use, demand, and resources) for Subsystems II-1&3 from each Hydrological Plan. Our water balance analysis did not include Subsystem II-2, located in the mountain range of Sierra Gorda-Zafarraya, because it constitutes an endorheic zone, meaning that its water resources are generated and used within the subsystem, mainly through exploitation of its aquifer. Also, the production of avocados and mangos in II-2 is minimal (Fig. 4A).

We masked LC classification rasters for 2009 and 2018 (103, 104) with various perimeter shapefiles to obtain aggregate surface area per LC category within Subsystems II-1&3 and within the perimeter of each groundwater body in SII-1&3. All data analyses and figures were done in R (123).

**Acknowledgments**

We are very grateful to all interviewees for their time and insights; and to the Junta de Andalucía, in particular SAIH Hidrosur and Sistema de Explotación Viñuela-Axarquía, for facilitating access to meteorological and hydrological data. We thank Sergio Vicente Serrano for the helpful explanation about handling zero precipitation values in SPI calculation.

**References**

1. A. Puy, *et al.*, The delusive accuracy of global irrigation water withdrawal estimates. *Nat. Commun.* **13**, 3183 (2022).
2. B. R. Scanlon, *et al.*, Global water resources and the role of groundwater in a resilient water future. *Nat. Rev. Earth Environ.* **4**, 87–101 (2023).
3. X. Huggins, *et al.*, Hotspots for social and ecological impacts from freshwater stress and storage loss. *Nat. Commun.* **13**, 439 (2022).
4. J. M. García-Ruiz, J. I. López-Moreno, S. M. Vicente-Serrano, T. Lasanta–Martínez, S. Beguería, Mediterranean water resources in a global change scenario. *Earth-Sci. Rev.* **105**, 121–139 (2011).
5. J. Rocha, *et al.*, Impacts of climate change on reservoir water availability, quality and irrigation needs in a water scarce Mediterranean region (southern Portugal). *Sci. Total Environ.* **736**, 139477 (2020).
6. P. H. Gleick, H. Cooley, Freshwater Scarcity. *Annu. Rev. Environ. Resour.* **46**, 319–348 (2021).
7. MAPA, Encuesta Sobre Superficies y Rendimientos de Cultivos, Resultados 2021. (2021). Available at: https://www.mapa.gob.es/es/estadistica/temas/estadisticas-agrarias/boletin2021_tcm30-623734.pdf [Accessed 27 May 2024].
8. FAOSTAT, FAOSTAT Data, Trade, Crops and livestock products, "Avocados" and "Mangos, guavas, and mangosteens", Import and Export quantities, by Region (Europe+), Year 2021. (2024). Available at: https://www.fao.org/faostat/en/#data/TCL [Accessed 4 July 2024].
9. R. Yus Ramos, Ó. Carrillo Romero, V. Fernández Camacho, M. Á. Torres Delgado, *La Burbuja de los Cultivos Subtropicales y el Colapso Hídrico de la Axarquía*, R. Yus Ramos, Ed. (Gabinete de Estudios de la Axarquía Vélez-Málaga (GENA), 2020).
10. A. R. Hurtado, E. Díaz-Cano, J. Berbel, The paradox of success: Water resources closure in Axarquia (southern Spain). *Sci. Total Environ.* **946**, 174318 (2024).
11. Axarquía Plus, El embalse de La Viñuela gana 3,9 hectómetros cúbicos tras las lluvias del invierno, aunque sigue la sequía grave. *AxarquiaPlus* (2024).
12. E. Cabezas, Vélez-Málaga y al menos otros cinco pueblos de la Axarquía cortan el agua por la noche. *D. Sur* (2023).
13. F. Extremera, Los grandes municipios abren pozos para paliar la peor sequía del siglo. *www.laopiniondemalaga.es* (2023).
14. A. Actis, La cosecha de mango se reduce un 80% en Málaga con acuíferos sobreexplotados y severas restricciones al consumo. *Política Online* (2023).
15. E. Cabezas, Ecologistas consideran que la gestión hídrica «dopa la burbuja de los subtropicales». *D. Sur* (2023). Available at: https://www.diariosur.es/axarquia/ecologistas-consideran-gestion-hidrica-dopa-burbuja-subtropicales-axarquia-20230306112642-nt.html [Accessed 27 May 2024].
16. P. Meyfroidt, Approaches and terminology for causal analysis in land systems science. *J. Land Use Sci.* **11**, 501–522 (2016).
17. N. R. Magliocca, *et al.*, Closing global knowledge gaps: Producing generalized knowledge from case studies of social-ecological systems. *Glob. Environ. Change* **50**, 1–14 (2018).

18. E. L. Tate, A. Gustard, "Drought Definition: A Hydrological Perspective" in *Drought and Drought Mitigation in Europe*, Advances in Natural and Technological Hazards Research., J. V. Vogt, F. Somma, Eds. (Springer Netherlands, 2000), pp. 23–48.
19. A. F. Van Loon, Hydrological drought explained. *WIREs Water* **2**, 359–392 (2015).
20. S. M. Vicente-Serrano, S. M. Quiring, M. Peña-Gallardo, S. Yuan, F. Domínguez-Castro, A review of environmental droughts: Increased risk under global warming? *Earth-Sci. Rev.* **201**, 102953 (2020).
21. L. Del Moral Ituarte, C. Giansante, Constraints to Drought Contingency Planning in Spain: The Hydraulic Paradigm and the Case of Seville. *J. Contingencies Crisis Manag.* **8**, 93–102 (2000).
22. D. Nortes Martínez, Structural change in Tagus River basin. Discussing the "'Eighties Effect'" in *VI Congress of the Spanish-Portuguese Association of Resource and Environmental Economics (AERNA)*, (2014).
23. A. Iglesias, M. Moneo, R. Garrote, F. Flores, "Drought and climate risks" in *Water Policy in Spain*, (CRC Press, 2009), p. Chapter 7.
24. IECA, Instituto de Estadística y Cartografía de Andalucía, Explotación de los Censos de Población y Vivienda del INE, Anual 2021. (2024). Available at: https://www.juntadeandalucia.es/institutodeestadisticaycartografia/badea/operaciones/consulta/anual/82862?CodOper=b3_3075&codConsulta=82862 [Accessed 14 June 2024].
25. IECA, Censo de Población 2001, Instituto de Estadística y Cartografía de Andalucía (IECA). (2024). Available at: https://www.juntadeandalucia.es/institutodeestadisticaycartografia/censos/censo2001/Tablas/index.htm [Accessed 10 July 2024].
26. E. Saiz, La batalla del agua en Andalucía: una guerra de cifras para eludir la deficiente gestión del campo. *El País* (2022).
27. F. J. Cantarero, A. J. Gallegos, J. Vías Martínez, D. Gumiel Muñoz, Accelerated change of land uses and repercussions on risks in the territory. The case of the extension of subtropical crops in the Axarquía of Malaga. in *Spanish Contribution to 34th International Geographical Congress*, (2020).
28. BOE, Real Decreto 594/1989, de 2 de junio, por el que se aprueba el Plan General de Transformación de la zona regable del Guaro (Málaga) (1989).
29. J. L. García-Aróstegui, J. Benavente, Procesos de intrusión-extrusión marina y propuestas de gestión integrada en acuíferos detríticos costeros mediterráneos: río Vélez (Málaga, España). *Bol. Geológico Min.* **118**, 671–682 (2007).
30. M. Pedro-Monzonís, P. Jiménez-Fernández, A. Solera, P. Jiménez-Gavilán, The use of AQUATOOL DSS applied to the System of Environmental-Economic Accounting for Water (SEEAW). *J. Hydrol.* **533**, 1–14 (2016).
31. Agrimensur, "Estudio de las infraestructuras hidráulicas para riego en la comarca de La Axarquía, Provincia de Málaga. Peticionario: Diputación Provincial de Málaga." (2017).
32. G. Rubio Galo, Axaragua y la Junta de Usuarios del Sur del Guaro acuerdan un plan de pagos por las aguas regeneradas. *Málaga Hoy* (2024).
33. A. González, Axaragua y Regantes Sur del Guaro acuerdan el pago por el uso de aguas regeneradas. *Opinión Málaga* (2024). Available at: https://www.laopiniondemalaga.es/axarquia/2024/04/08/axaragua-regantes-sur-guaro-acuerdan-pago-aguas-regeneradas-100789282.html [Accessed 31 May 2024].
34. JdA, La Junta multiplica por cuatro el volumen de aguas regeneradas de Andalucía con sus obras de hídricas. *Junta Andal.* (2024). Available at: https://www.juntadeandalucia.es/organismos/agriculturapescaaguaydesarrollorural/servicios/actualidad/noticias/detalle/495379.html [Accessed 18 July 2024].
35. JdA, El Gobierno de Andalucía invierte 50 M€ en los terciarios y obras de abastecimiento de La Axarquía. *Junta Andal.* (2023). Available at: https://www.juntadeandalucia.es/organismos/agriculturapescaaguaydesarrollorural/servicios/actualidad/noticias/detalle/448807.html [Accessed 18 July 2024].

36. M.-T. Sánchez-Martínez, M. Salas-Velasco, N. Rodríguez-Ferrero, Who Manages Spain's Water Resources? The Political and Administrative Division of Water Management. *Int. J. Water Resour. Dev.* **28**, 27–42 (2012).
37. S. Isselhorst, J. Berking, B. Schütt, "Irrigation Communities and Agricultural Water Management in Andalusia. A Special Focus on the Vega of Vélez Blanco" in *Water Management in Ancient Civilizations*, Berlin Studies of the Ancient World., J. Berking, Ed. (Edition Topoi, 2018).
38. E. Custodio, Consideraciones sobre el pasado, presente y futuro de las aguas subterráneas en España. *Ing. Agua* **26**, 1–17 (2022).
39. N. Hernández-Mora, L. Martínez-Cortina, M. R. Llamas Madurga, E. Custodio Gimena, Groundwater issues in southern EU Member States: Spain country report in *Working Group Appointed by the European Academies of Sciences Advisory Council (EASAC)*, (2007).
40. MITECO, ¿Qué es el Registro de Aguas? *Minist. Para Transic. Ecológica El Reto Demográfico* (2024). Available at: https://www.miteco.gob.es/es/agua/temas/concesiones-y-autorizaciones/uso-privativo-del-agua-registro-del-aguas/registros-aguas-desde-1986/que_es_registro_aguas.html [Accessed 31 May 2024].
41. M. M. Aldaya, *et al.*, An academic analysis with recommendations for water management and planning at the basin scale: A review of water planning in the Segura River Basin. *Sci. Total Environ.* **662**, 755–768 (2019).
42. BOE, Boletín Oficial del Estado, Orden ARM/2656/2008, de 10 de septiembre, por la que se aprueba la instrucción de planificación hidrológica, Ministerio de Medio Ambiente, y Medio Rural y Marino (2008).
43. M. R. Llamas, E. Custodio, A. De La Hera, J. M. Fornés, Groundwater in Spain: increasing role, evolution, present and future. *Environ. Earth Sci.* **73**, 2567–2578 (2015).
44. F. López-Vera, Groundwater: The Invisible Resource. *Int. J. Water Resour. Dev.* **28**, 141–150 (2012).
45. WWF, "Illegal water use in Spain: Causes, effects and solutions" (2006).
46. M. R. Llamas, A. Garrido, "Lessons from intensive groundwater use in Spain: Economic and social benefits and conflicts" in *The Agricultural Groundwater Revolution: Opportunities and Threats to Development*, 1st Ed., M. Giordano, K. G. Villholth, Eds. (CABI, 2007).
47. E. Lopez-Gunn, The Role of Collective Action in Water Governance: A Comparative Study of Groundwater User Associations in La Mancha Aquifers in Spain. *Water Int.* **28**, 367–378 (2003).
48. OECD, Land-use Planning Systems in the OECD: Country Fact Sheets: Spain. *Oecd-Ilibraryorg* (2017). Available at: https://read.oecd-ilibrary.org/urban-rural-and-regional-development/land-use-planning-systems-in-the-oecd_9789264268579-en [Accessed 31 May 2024].
49. M. Benavent F. de Córdoba, "Los Planes de Ordenación del Territorio en España: De la instrumentación a la gestión" in *Agua, Territorio y Paisaje*, (Fundicot, Asociación Interprofesional de Ordenación del Territorio, 2009), pp. 143–158.
50. J. Pérez-Sánchez, J. Senent-Aparicio, Integrated water resources management on a local scale: a challenge for the user community—a case study in Southern Spain. *Environ. Earth Sci.* **74**, 6097–6109 (2015).
51. A. L. Grindlay, Implementation of the European Water Framework Directive: Integration of hydrological and regional planning at the Segura River Basin, southeast Spain. *Land Use Policy* (2011).
52. A. Garrido, P. Martínez-Santos, M. R. Llamas, Groundwater irrigation and its implications for water policy in semiarid countries: the Spanish experience. *Hydrogeol. J.* **14**, 340–349 (2006).
53. JdA, Canon de regulación del agua del Embalse de la Viñuela, Año 2024, Junta de Andalucía, Consejería de Agricultura, Pesca, Agua y Desarrollo Rural (2024).
54. JdA, Plan Hidrológico 2022-2027, Demarcación Hidrográfica de las Cuencas Mediterráneas Andaluzas (DHCMA) - Anejo IX Recuperación de Costes (2023).

55. ASAJA, Balance Técnicos 2021. (2022).
56. G. Moreno-Ortega, C. Pliego, D. Sarmiento, A. Barceló, E. Martínez-Ferri, Yield and fruit quality of avocado trees under different regimes of water supply in the subtropical coast of Spain. *Agric. Water Manag.* **221**, 192–201 (2019).
57. V. H. Durán Zuazo, C. R. Rodríguez Pleguezuelo, B. Gálvez Ruiz, S. Gutiérrez Gordillo, I. F. García-Tejero, Water use and fruit yield of mango ( *Mangifera indica* L.) grown in a subtropical Mediterranean climate. *Int. J. Fruit Sci.* **19**, 136–150 (2019).
58. BOJA, Orden de 22 de mayo de 2024, por la que se declaran los cambios de estado y las medidas a adoptar debido al estado de sequía en las Cuencas Mediterráneas Andaluzas. Boletín Oficial de la Junta de Andalucía (2024).
59. JdA, Estadística de superficies y producciones de los cultivos agrícolas en Andalucía (Avances de superficies y producciones). - Actividades estadísticas y cartográficas, Junta de Andalucía. (2023). Available at: https://www.juntadeandalucia.es/organismos/agriculturapescaaguaydesarrollorural/servicios/estadistica-cartografia/actividad/detalle/175051.html [Accessed 31 May 2024].
60. JdA, Plan Hidrológico de las Cuencas Mediterráneas Andaluzas 2009-2015 - ANEJO III. Usos y Demandas. Junta de Andalucía. *Portal Ambient. Andal.* (2011). Available at: https://www.juntadeandalucia.es/medioambiente/portal/documents/20151/1779232/ANEJO_III.pdf/0cca0244-b608-48c5-a822-2ddca14ccaad [Accessed 31 May 2024].
61. JdA, Plan Hidrológico 2022-2027, Demarcación Hidrográfica de las Cuencas Mediterráneas Andaluzas (DHCMA) - Anejo VI: Asignación y Reserva de Recursos a Usos (2023).
62. JdA, "Informe de Escasez y Sequía, SAIH HIDROSUR, Consejería de Agricultura, Pesca, Agua y Desarrollo Rural, Direcci±n General de Infraestructuras del Agua, Junta de Andalucía" (2024).
63. M. Cortés, Alerta en el pantano de La Viñuela, con los niveles más bajos de los últimos 13 años. *Málaga Hoy* (2021). Available at: https://www.malagahoy.es/malaga/pantano-Vinuela-niveles_0_1630937190.html [Accessed 26 July 2024].
64. M. A. Palmer, *et al.*, Climate change and the world's river basins: anticipating management options. *Front. Ecol. Environ.* **6**, 81–89 (2008).
65. R. R. Brown, M. A. Farrelly, Delivering sustainable urban water management: a review of the hurdles we face. *Water Sci. Technol.* **59**, 839–846 (2009).
66. BOJA, Acuerdo de 4 de mayo de 2021, del Consejo de Gobierno, por el que se aprueba el Plan Especial de Actuación en situación de alerta y eventual sequía para la Demarcación Hidrográfica de las Cuencas Mediterráneas Andaluzas. (2021). Available at: https://www.juntadeandalucia.es/boja/2021/86/37.html [Accessed 16 July 2024].
67. BOJA, Decreto 178/2021, de 15 de junio, por el que se regulan los indicadores de sequía hidrológica y las medidas excepcionales para la gestión de los recursos hídricos en las Demarcaciones Hidrográficas Intracomunitarias de Andalucía. (2021). Available at: https://www.juntadeandalucia.es/boja/2021/116/24 [Accessed 13 June 2024].
68. JdA, Plan Hidrológico de las Cuencas Mediterráneas Andaluzas 2009-2015 - Junta de Andalucía. *Portal Ambient. Andal.* (2011). Available at: https://www.juntadeandalucia.es/medioambiente/portal/areas-tematicas/agua/planificacion-hidrologica/2009-2015/cuencas-mediterraneas-andaluzas [Accessed 31 May 2024].
69. JdA, Las Cuencas Mediterráneas Andaluzas, excepto Granada, en situación de excepcional sequía. *Junta Andal.* (2022). Available at: https://www.juntadeandalucia.es/presidencia/portavoz/tierraymar/170079/Sequia/AguaRestricciones/CuencasMediterraneasAndaluzas/InfraestructurasHidraulicas [Accessed 17 July 2024].
70. BOJA, Decreto-ley 2/2024, de 29 de enero, por el que se aprueban medidas adicionales para paliar los efectos producidos por la situación de excepcional sequía a los usuarios de las demarcaciones hidrográficas intracomunitarias de Andalucía y se adoptan medidas urgentes, administrativas y fiscales, de apoyo al sector agrario. Boletín Oficial de la Junta de Andalucía. (2024).

71. V. Junquera, *et al.*, Crop booms as regime shifts. *R. Soc. Open Sci.* **11**, 231571 (2024).
72. C. Folke, Resilience: The emergence of a perspective for social–ecological systems analyses. *Glob. Environ. Change* **16**, 253–267 (2006).
73. N. Ramankutty, O. T. Coomes, Land-use regime shifts: an analytical framework and agenda for future land-use research. *Ecol. Soc.* **21**, art1 (2016).
74. M. Milkoreit, *et al.*, Defining tipping points for social-ecological systems scholarship—an interdisciplinary literature review. *Environ. Res. Lett.* **13**, 033005 (2018).
75. E. W. Tekwa, E. P. Fenichel, S. A. Levin, M. L. Pinsky, Path-dependent institutions drive alternative stable states in conservation. *Proc. Natl. Acad. Sci.* **116**, 689–694 (2019).
76. V. Dakos, S. R. Carpenter, E. H. van Nes, M. Scheffer, Resilience indicators: prospects and limitations for early warnings of regime shifts. *Philos. Trans. R. Soc. B Biol. Sci.* **370**, 20130263 (2015).
77. M. W. Macy, M. Ma, D. R. Tabin, J. Gao, B. K. Szymanski, Polarization and tipping points. *Proc. Natl. Acad. Sci.* **118**, e2102144118 (2021).
78. R. Pedreschi, *et al.*, Short vs. Long-Distance Avocado Supply Chains: Life Cycle Assessment Impact Associated to Transport and Effect of Fruit Origin and Supply Conditions Chain on Primary and Secondary Metabolites. *Foods* **11**, 1807 (2022).
79. JdA, Plan de Ordenación del Territorio de la Costa del Sol Oriental-Axarquía (Málaga) - Memoria Informativa (2006).
80. J. Rodrigo-Comino, A. Caballero-Calvo, L. Salvati, J. M. Senciales-González, Sostenibilidad de los cultivos subtropicales: claves para el manejo del suelo, el uso agrícola y la Ordenación del Territorio. *Cuad. Geográficos* **61**, 150–167 (2022).
81. J. Hernández Bedolla, A. Solera, J. Paredes Arquiola, C. X. Roblero Escobar, Análisis del cambio en las aportaciones hidrológicas en la cuenca del río Júcar a partir de 1980 y sus causas. *Ing. Agua* **23**, 141 (2019).
82. J. Molina, J. A. Sillero-Medina, J. D. Ruiz-Sinoga, Modeling the Rainfall Exploitation of the Reservoirs in Malaga Province, Spain. *Air Soil Water Res.* **16**, 11786221231185104 (2023).
83. Tragsatec, TRABAJOS DE APOYO TÉCNICO PARA LA DETERMINACIÓN DE LAS SUPERFICIES EN REGADÍO EN LAS DEMARCACIONES HIDROGRÁFICAS INTRACOMUNITARIAS ANDALUZAS EMPLEANDO TÉCNICAS DE TELEDETECCIÓN. SEGUIMIENTO DEL REGADÍO EN LA DEMARCACIÓN HIDROGRÁFICA DE LAS CUENCAS MEDITERRÁNEAS ANDALUZAS (DHCMA)-SISTEMAS 2 y 3. Deposited August 2019.
84. E. Cabezas, Las comunidades de regantes investigadas en la Axarquía guardan silencio. *D. Sur* (2023). Available at: https://www.diariosur.es/sucesos/comunidades-regantes-investigadas-axarquia-guardan-silencio-operacion-guardia-civil-20230509211050-nt.html [Accessed 1 June 2024].
85. J. Cano, El Seprona detecta 250 pozos ilegales en la operación contra el saqueo de agua en la Axarquía. *D. Sur* (2023). Available at: https://www.diariosur.es/sucesos/seprona-detecta-250-pozos-ilegales-operacion-saqueo-agua-axarquia-malaga-20230508010229-nt.html [Accessed 1 June 2024].
86. E. Cabezas, El Seprona cifra en 25 hectómetros el expolio de los regadíos ilegales de subtropicales en la Axarquía. *D. Sur* (2024). Available at: https://www.diariosur.es/sucesos/pozos-ilegales-regadios-20240205131410-nt.html [Accessed 25 July 2024].
87. JdA, Plan de Ordenación del Territorio de la Costa del Sol Oriental-Axarquía (Málaga) - Normativa (2006).
88. JdA, Plan de Ordenación del Territorio de la Costa del Sol Oriental-Axarquía (Málaga) - Memoria de Ordenación (2006).
89. BOE, Real Decreto 329/2002, de 5 de abril, por el que se aprueba el Plan Nacional de Regadíos. Boletín Oficial del Estado. Ministerio de Agricultura, Pesca y Alimentación (2002).
90. J. L. Molina, *et al.*, Aquifers Overexploitation in SE Spain: A Proposal for the Integrated Analysis of Water Management. *Water Resour. Manag.* **23**, 2737–2760 (2009).

91. JdA, Costes producción: Aguacate 2011/12. Observatorio de Precios y Mercados. Consejería de Agricultura, Pesca y Desarrollo Rural. Junta de Andalucía. (2024). Available at: https://www.juntadeandalucia.es/agriculturaypesca/observatorio/servlet/FrontController?action=RecordContent&table=11200&element=880575&subsector=& [Accessed 1 August 2024].
92. JdA, Average production costs. Crop year 2012. Mango. Observatorio de Precios y Mercados. Consejería de Agricultura, Pesca y Desarrollo Rural. Junta de Andalucía. (2024). Available at: https://www.juntadeandalucia.es/agriculturaypesca/observatorio/servlet/FrontController?action=RecordContent&table=11200&element=1038830&subsector=& [Accessed 1 August 2024].
93. J. C. Leyva, S. Sayadi, Economic valuation of water and willingness to pay analysis in tropical fruit production in South-Eastern Spain. *Span. J. Agric. Res.* **3**, 25–33 (2005).
94. N. L. Engle, O. R. Johns, M. C. Lemos, D. R. Nelson, Integrated and Adaptive Management of Water Resources: Tensions, Legacies, and the Next Best Thing. *Ecol. Soc.* **16** (2011).
95. E. Ostrom, Background on the Institutional Analysis and Development Framework. *Policy Stud. J.* **39**, 7–27 (2011).
96. E. M. Bruno, K. Jessoe, Designing water markets for climate change adaptation. *Nat. Clim. Change* **14**, 331–339 (2024).
97. R. J. P. Schmitt, L. Rosa, G. C. Daily, Global expansion of sustainable irrigation limited by water storage. *Proc. Natl. Acad. Sci.* **119**, e2214291119 (2022).
98. P.-W. Liu, *et al.*, Groundwater depletion in California's Central Valley accelerates during megadrought. *Nat. Commun.* **13**, 7825 (2022).
99. M. Kang, D. Perrone, Z. Wang, S. Jasechko, M. M. Rohde, Base of fresh water, groundwater salinity, and well distribution across California. *Proc. Natl. Acad. Sci.* **117**, 32302–32307 (2020).
100. R. Q. Grafton, *et al.*, Global insights into water resources, climate change and governance. *Nat. Clim. Change* **3**, 315–321 (2013).
101. R. Noori, *et al.*, Anthropogenic depletion of Iran's aquifers. *Proc. Natl. Acad. Sci.* **118**, e2024221118 (2021).
102. C. J. Bauer, Water Conflicts and Entrenched Governance Problems in Chile's Market Model. **8** (2015).
103. Tragsatec, Electronic dataset: REGADIO_2009/CLAS_DHCMA_S2_S3_2009.tif, Archivo ráster (en formato *.tif) que contiene la información correspondiente a la clasificación de los cultivos en regadío del año 2009. Deposited August 2019.
104. Tragsatec, Electronic dataset: REGADIO_2018/CLAS_ DHCMA_S2_S3_2018.tif: Archivo ráster (en formato *.tif) que contiene la información correspondiente a la clasificación de los cultivos en regadío del año 2018. Deposited August 2019.
105. Tragsatec, Electronic dataset (shapefiles): COMUNIDADES_REGANTES/Regadios_existentes_CMA_S2_S3_desagregadoSIGPAC 18: contiene los vectoriales en formato SHAPE de las comunidades de regantes desagregadas. Deposited August 2019.
106. Tragsatec, Electronic dataset (shapefiles): COMUNIDADES_REGANTES/Recintos_SIGPAC_reg_DHCMA_S2_S3_2018_complementarios: contiene los vectoriales en formato SHAPE de los recintos de regadío complementarios de 2018 con los datos de superficie de regadío incorporados. Deposited August 2019.
107. DIVA-GIS, DIVA-GIS Free Spatial Data - Spain. Deposited 2024.
108. REDIAM, Catálogo REDIAM (Red de Información Ambiental de Andalucía). Cartografía derivada de los Planes Hidrológicos de las Demarcaciones intracomunitarias andaluzas (para Agua0). (2019). Available at: https://portalrediam.cica.es/geonetwork/srv/eng/catalog.search#/metadata/3cd10003-3962-4a8a-a5b3-58e1782a2a3b [Accessed 10 June 2024].

109. IECA, Datos Espaciales de Referencia de Andalucía (DERA), Instituto de Estadística y Cartografía de Andalucía - 1. Relieve. (2024). Available at: https://www.juntadeandalucia.es/institutodeestadisticaycartografia/dega/datos-espaciales-de-referencia-de-andalucia-dera/descarga-de-informacion [Accessed 31 May 2024].
110. JdA, Plan Hidrológico Cuenca del Sur, 1998 - Junta de Andalucía. *Portal Ambient. Andal.* (1998). Available at: https://www.juntadeandalucia.es/medioambiente/portal/areas-tematicas/agua/planificacion-hidrologica/1998/cuenca-del-sur [Accessed 31 May 2024].
111. JdA, Plan Hidrológico de las Cuencas Mediterráneas 2015-2021 - Junta de Andalucía. *Portal Ambient. Andal.* (2016). Available at: https://www.juntadeandalucia.es/medioambiente/portal/areas-tematicas/agua/planificacion-hidrologica/2015-2021/cuencas-mediterraneas [Accessed 31 May 2024].
112. BOE, Boletín Oficial del Estado (2019).
113. T. B. McKee, N. J. Doesken, J. Kleist, THE RELATIONSHIP OF DROUGHT FREQUENCY AND DURATION TO TIME SCALES. *Eighth Conf. Appl. Climatol. 17-22 January 1993 Anaheim Calif.* (1993).
114. V. Gumus, Evaluating the effect of the SPI and SPEI methods on drought monitoring over Turkey. *J. Hydrol.* **626**, 130386 (2023).
115. H. Aksoy, *et al.*, SPI-based Drought Severity-Duration-Frequency Analysis. (2018).
116. H. Wu, M. D. Svoboda, M. J. Hayes, D. A. Wilhite, F. Wen, Appropriate application of the standardized precipitation index in arid locations and dry seasons. *Int. J. Climatol.* **27**, 65–79 (2007).
117. D. G. C. Kirono, V. Round, C. Heady, F. H. S. Chiew, S. Osbrough, Drought projections for Australia: Updated results and analysis of model simulations. *Weather Clim. Extrem.* **30**, 100280 (2020).
118. S. Beguería, S. M. Vicente-Serrano, SPEI: Calculation of the Standardized Precipitation-Evapotranspiration Index version 1.8.1 from CRAN. (2023). Deposited 7 March 2023.
119. Z. Hao, R: Statistical Modeling and Assessment of Drought, Documentation for package 'drought' version 1.2. (2024). Available at: https://search.r-project.org/CRAN/refmans/drought/html/00Index.html [Accessed 31 May 2024].
120. J. Zhou, M. J. Deitch, S. Grunwald, E. Screaton, Do the Mann-Kendall test and Theil-Sen slope fail to inform trend significance and magnitude in hydrology? *Hydrol. Sci. J.* **68**, 1241–1249 (2023).
121. A. I. McLeod, Kendall: Kendall Rank Correlation and Mann-Kendall Trend Test. (2022). Deposited 20 March 2022.
122. J. Raymaekers, robslopes: Fast Algorithms for Robust Slopes. (2023). Deposited 27 April 2023.
123. R Core Team, R: A Language and Environment for Statistical Computing. (2023). Deposited 2023.

**Figures and Tables**

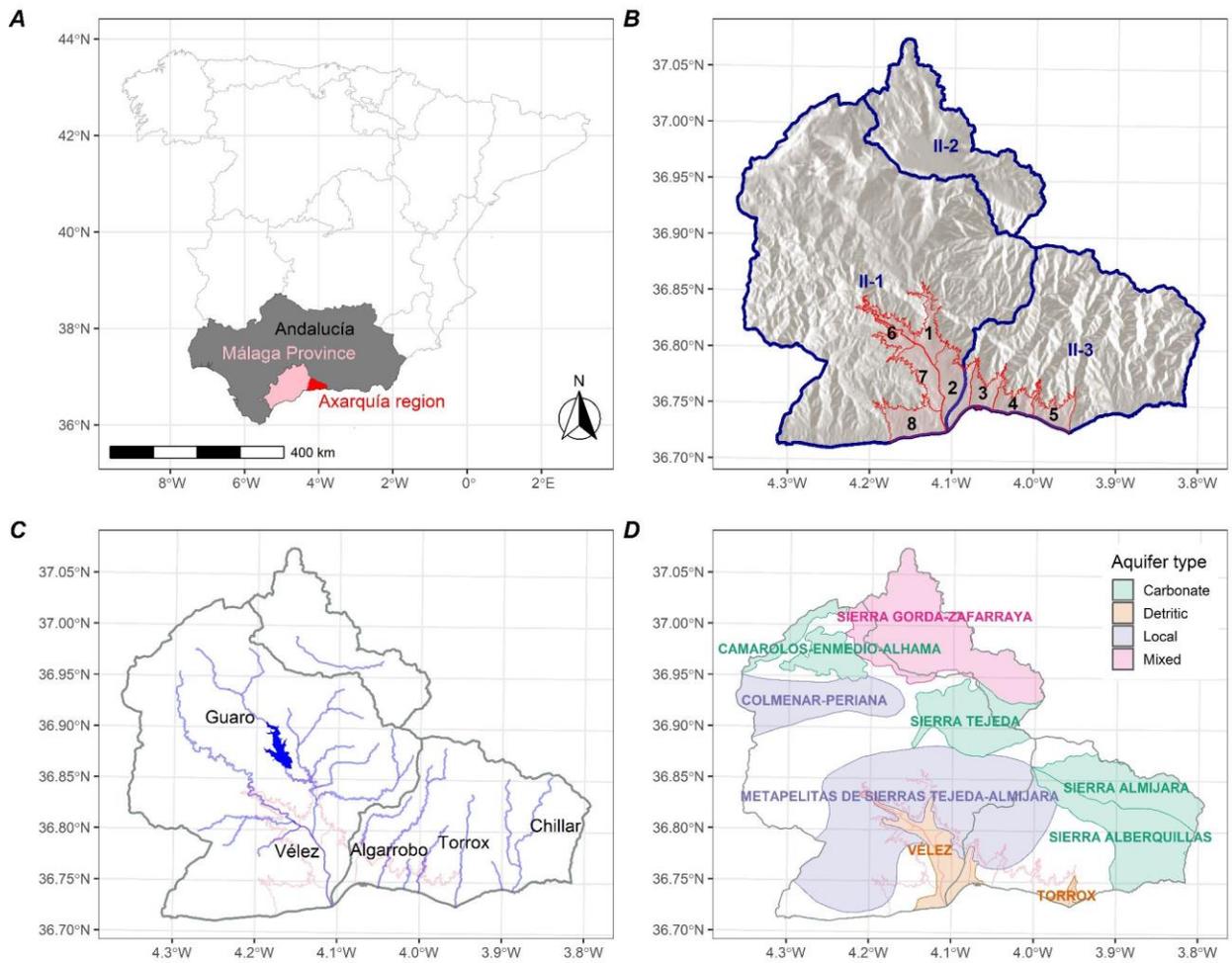

**Figure 1.** Maps of Axarquía study region. A) Location of Axarquía within mainland Spain. B) Axarquía region (System II), comprising Subsystems II-1,2,3. Plan Guaro (PG) boundary and sectors (1-5: left-side; 6-8: right-side) (red). C) Viñuela reservoir (blue) and main rivers. D) Groundwater bodies and types.

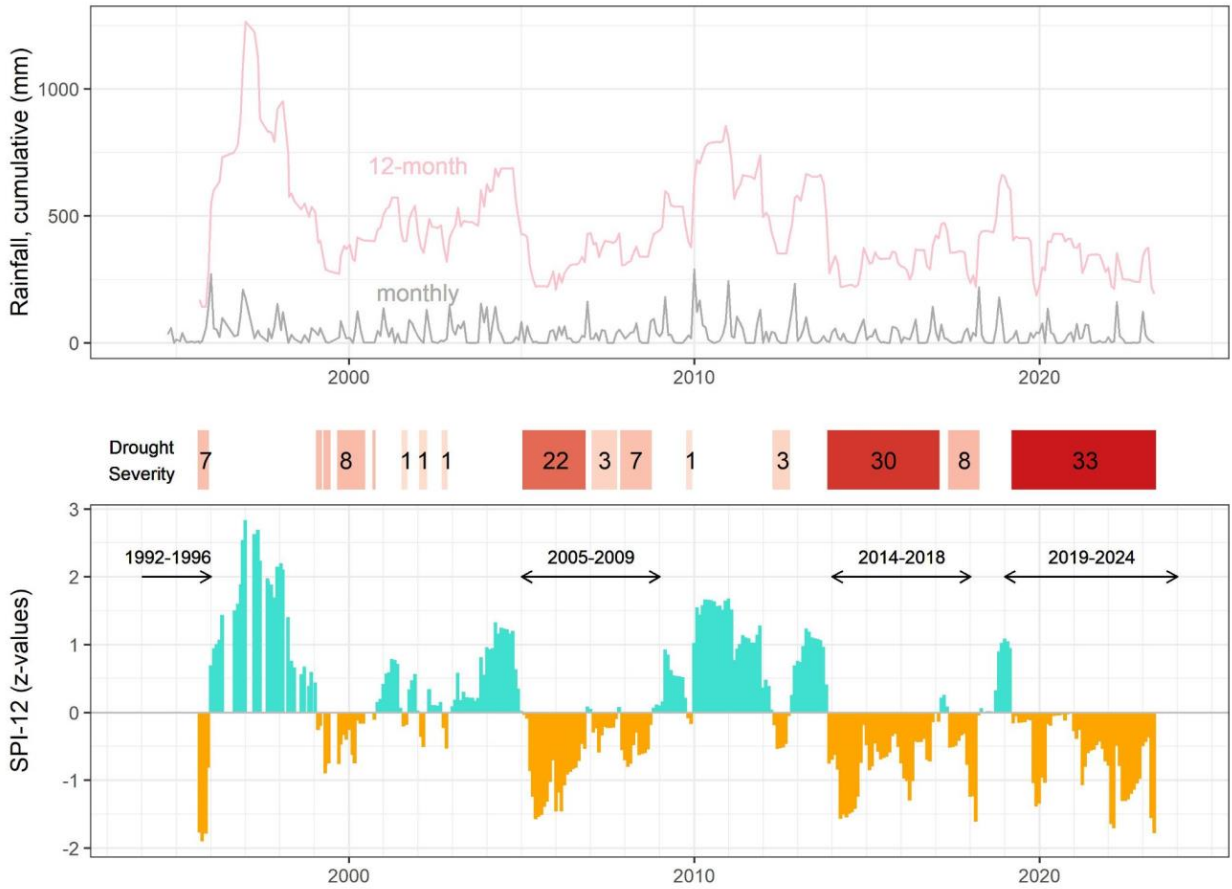

**Figure 2.** Cumulative rainfall, drought severity index for each drought event, and 12-month Standard Precipitation Index (SPI-12), including major droughts (arrows).

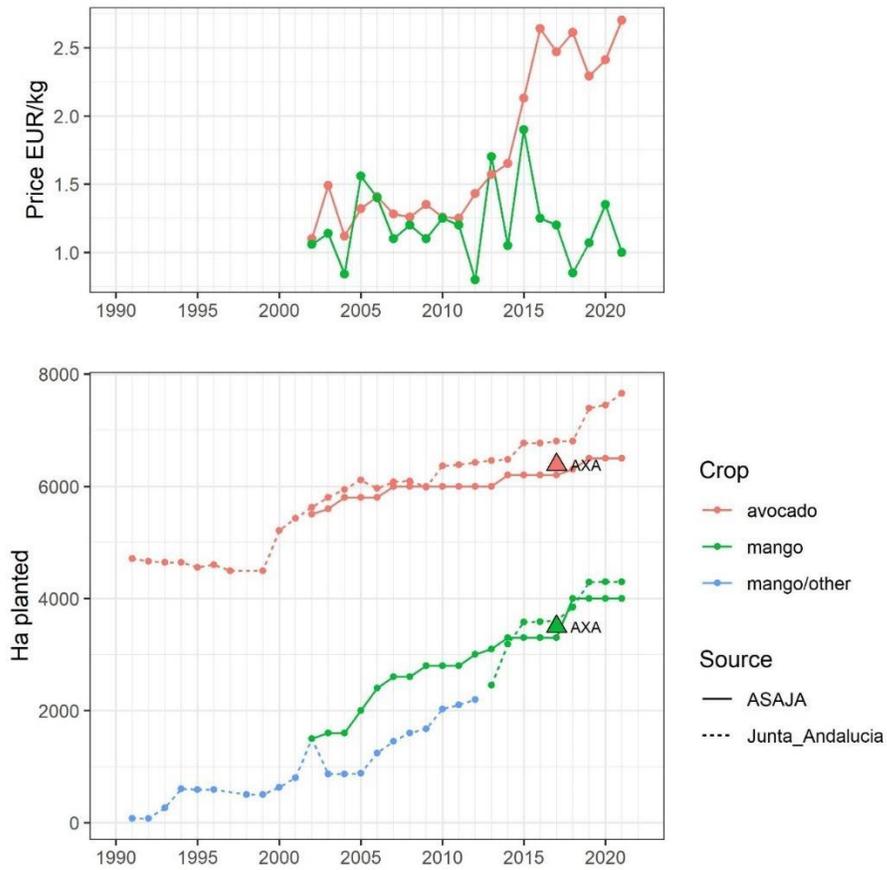

**Figure 3.** Evolution of avocado and mango prices (top) and planted area (bottom) in Málaga Province. Sources: ASAJA (ASAJA 2022), Junta_Andalucia (59). AXA = Values for Axarquía only (Yus Ramos et al., 2020, p. 476).

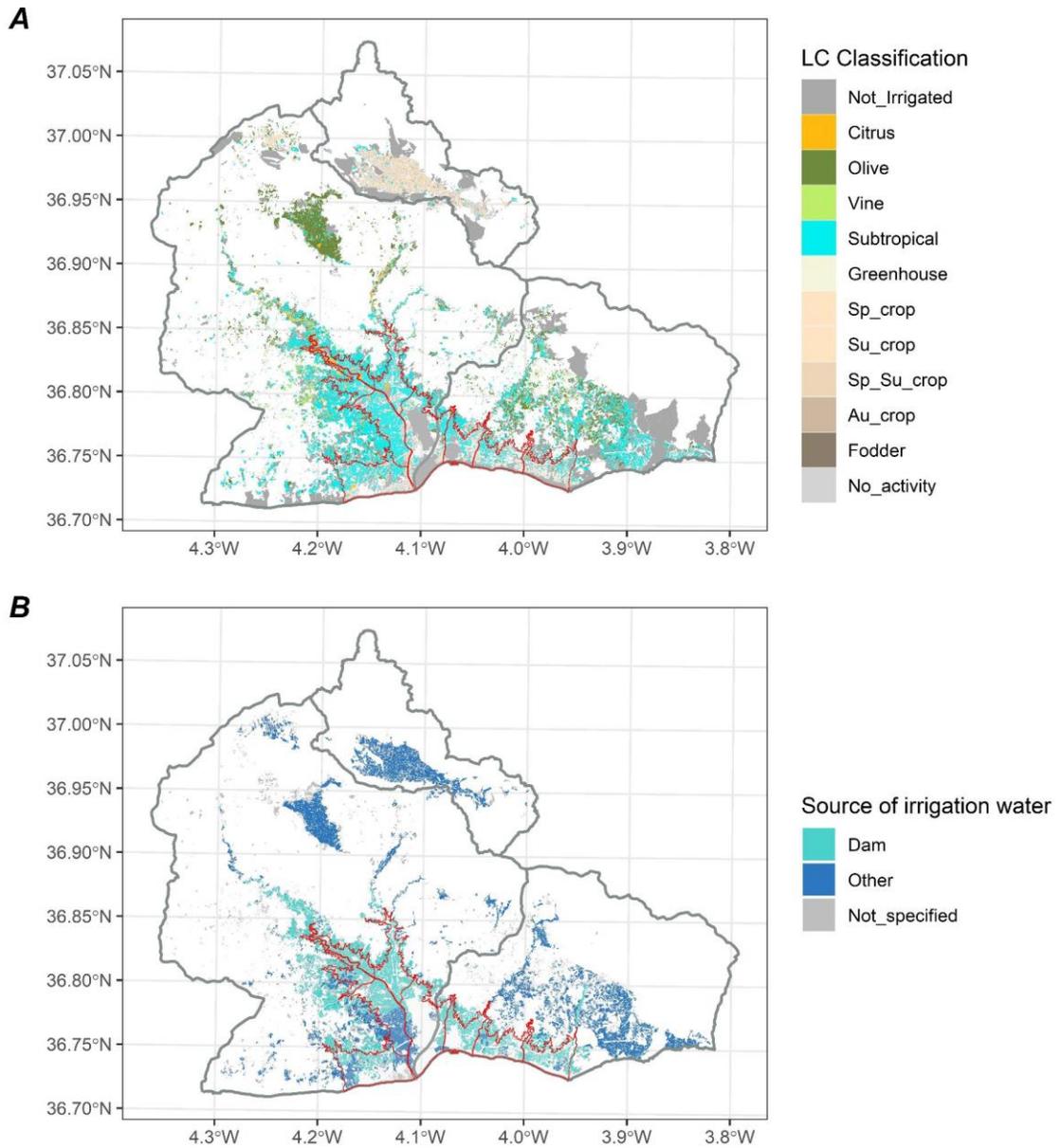

**Figure 4.** Extent of subtropical fruit plantations and other land cover (LC) categories in 2018 (Source: 104). B) Areas irrigated with dam water in 2018 clearly exceed the Plan Guaro (PG) perimeter (red line) (Source: 105, 106).

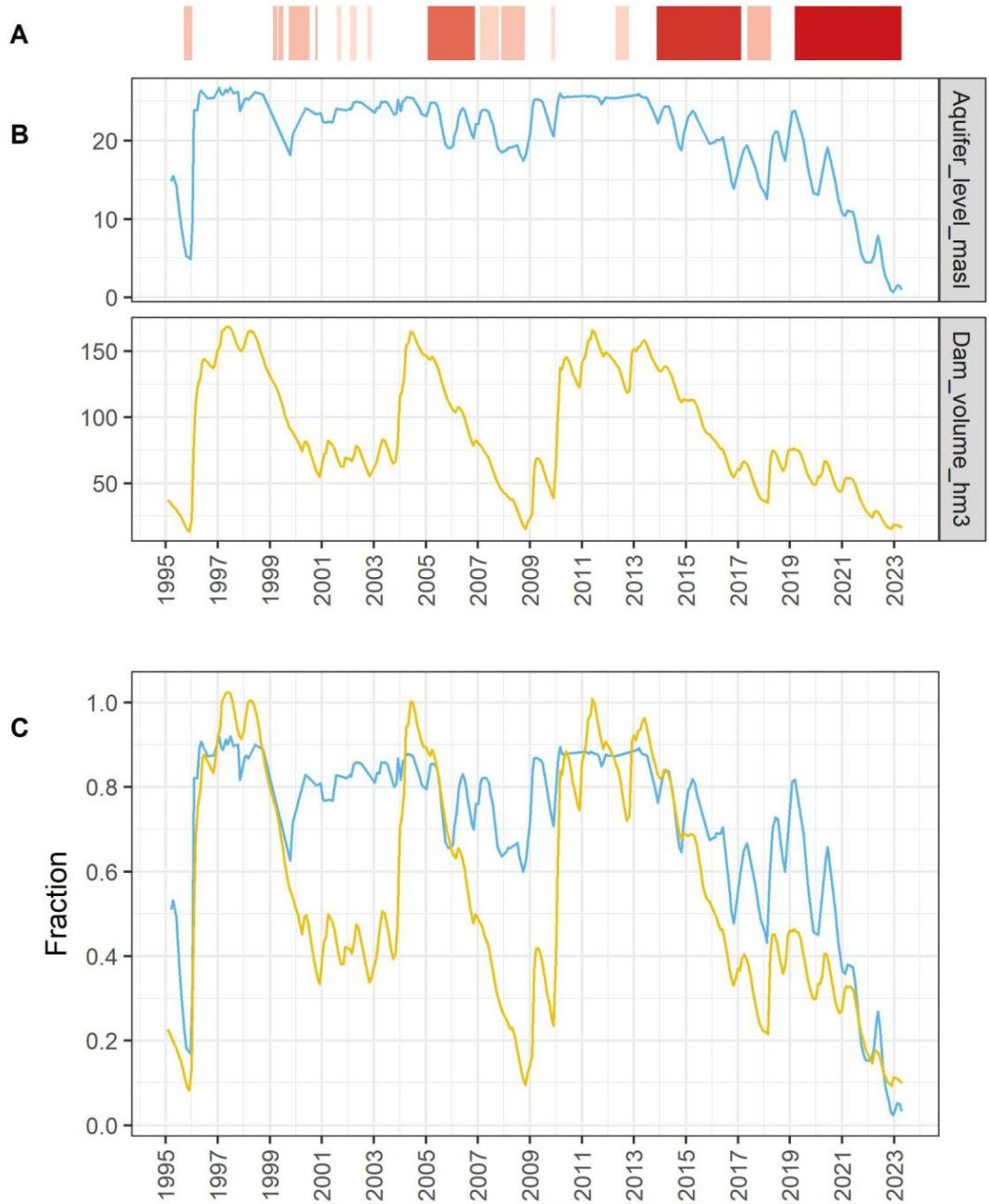

**Figure 5.** (A) Drought severity (From Fig. 2). (B) Vélez aquifer level measured by piezometer 013-S (blue) and dam volume (yellow). (C) Vélez aquifer level fraction and dam volume fraction.

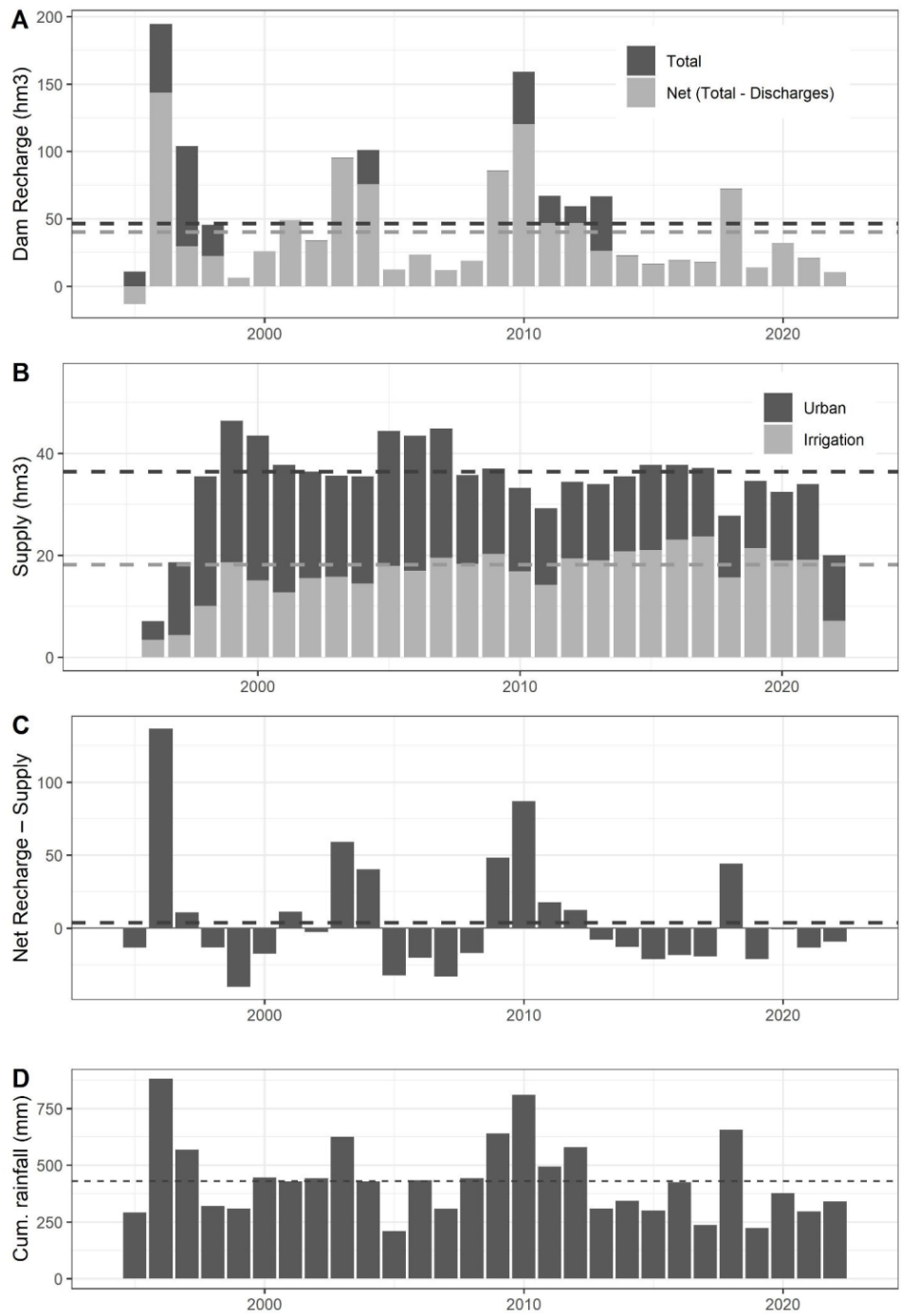

**Figure 6:** (A)-(C): Viñuela dam operating parameters (in hm3): (A) Total Recharge (from precipitation, runoff and other sources, and subtracting evaporation); Net Recharge = Total Recharge minus Discharges (water not used and discharged for environmental or safety purposes). (B) Dam water Supply for Irrigation and Urban uses. (C) Net Recharge minus Supply (dam water balance). (D) Cumulative annual rainfall. Note: Dotted lines indicate the mean for the period 2000-2021 in all graphs.

**Table 1.** Water balance (use and demand) in Subsystems II-1&3 in all four Hydrological Plans (PH). All values in hm3/year unless otherwise noted. Source: own elaboration.

| | | Water use (estimate) | | | | | Water demand (theoretical) | | | | | | | |
|---|---|---|---|---|---|---|---|---|---|---|---|---|---|---|
| Hydrological Plan | Subsystem | Dam | Surface | Groundwater | Regen. Wastewater | Total Use (U) | Urban | Irrigation | Livestock | Golf | Total Demand (D) | Difference (R-D) | Irrigated ha | Water use m3/ha |
| PH 1998 | SII-1 | 33 | 16 | 32 | 0.0 | 79 | 15 | 68 | 0 | 1 | 84 | -4.8 | 9655 | 6415 |
| | SII-3 | 0 | 10 | 14 | 0.6 | 26 | 7 | 26 | 0 | 0 | 32 | -5.7 | 3650 | 6294 |
| PH 2009-2015 | SII-1 | 37 | 4.8 | 20 | 0.2 | 55 | 15 | 42 | 0.2 | 0.4 | 58 | -2.6 | 8976 | 4295 |
| | SII-3 | 0 | 4.4 | 11 | 0.0 | 22 | 7.7 | 21 | 0.0 | 0.4 | 30 | -7.1 | 4349 | 3308 |
| PH 2015-2021 | SII-1 | 37 | 4.8 | 19 | 0.2 | 55 | 15 | 45 | 0.2 | 0.4 | 60 | -5.1 | 8976 | 4295 |
| | SII-3 | 0 | 4.4 | 12 | 0.0 | 23 | 8.1 | 24 | 0.0 | 0.4 | 32 | -9.3 | 4349 | 3308 |
| PH 2021-2027 | SII-1&3 | 32 | 15 | 45 | 0.3 | 93 | 24 | 78 | 0.2 | 0.8 | 103 | -11 | 15034 | 5214 |
| **Total Subsystems II-1&3** | | | | | | | | | | | | | | |
| PH 1998 | SII-1&3 | 33 | 26 | 47 | 0.6 | 106 | 22 | 93 | 0 | 1 | 116 | -10 | 13305 | 6382 |
| PH 2009-2015 | SII-1&3 | 37 | 9.2 | 31 | 0.2 | 78 | 23 | 64 | 0.2 | 0.8 | 88 | -9.7 | 13325 | 3973 |
| PH 2015-2021 | SII-1&3 | 37 | 9.2 | 31 | 0.2 | 78 | 23 | 68 | 0.2 | 0.8 | 93 | -14 | 13325 | 3973 |
| PH 2021-2027 | SII-1&3 | 32 | 15 | 45 | 0.3 | 93 | 24 | 78 | 0.2 | 0.8 | 103 | -11 | 15034 | 5214 |

**Table 2.** Information in the Hydrological Plans (PH) for the three aquifers in SII-1&3, including groundwater (GW) use by sector; groundwater resources, sustainable extraction rates, and aquifer exploitation index; and groundwater permitting status. All units are in hm3/year unless otherwise noted. Note: Torrox and Metapelitas were not included in PH 1998. Source: own elaboration.

| Hydrological Plan | Aquifer | GW use by sector ||||| GW resources ||| Permits |||
|---|---|---|---|---|---|---|---|---|---|---|---|---|
| | | Use, Irrigation | Use, Urban | Use, Golf | Use, Industry | Total Use | Resource, total | Resource, sustainable (RS) | Exploitation Index (Total Use / RS) (unitless) | Registered (Reg) | Under Review (UR) | Reg + UR |
| PH 1998 | Velez | 27 | 0 | 0.43 | 0 | 27 | 33 | 28 | 0.96 | no data |||
| PH 2009-2015 | | 12 | 0.07 | 0.41 | 0 | 12 | 24 | 13 | 0.94 | 8.6 | 11 | 20 |
| PH 2015-2021 | | 12 | 0.3 | 0.41 | 0 | 12 | 24 | 14 | 0.87 | 8.6 | 11 | 20 |
| PH 2021-2027 | | 18 | 4.2 | 0.41 | 0.18 | 23 | 22 | 14 | 1.6 | 14 | 5.76 | 20 |
| PH 2009-2015 | Torrox | 0.27 | 0 | 0 | 0 | 0.27 | 0.7 | 0.42 | 0.64 | 0.27 | 0.39 | 0.66 |
| PH 2015-2021 | | 0.27 | 0 | 0 | 0 | 0.27 | 0.7 | 0.42 | 0.64 | 0.27 | 0.39 | 0.66 |
| PH 2021-2027 | | 0.71 | 0.13 | 0 | 0 | 0.84 | 1 | 0.70 | 1.2 | 0.79 | 0.02 | 1 |
| PH 2009-2015 | Metapelitas | 3.6 | 0.23 | 0.24 | 0 | 4.0 | 6.2 | 5.0 | 0.81 | 8.0 | 7.3 | 15 |
| PH 2015-2021 | | 3.6 | 0.23 | 0.24 | 0 | 4.0 | 6.2 | 5.0 | 0.81 | 8.0 | 7.3 | 15 |
| PH 2021-2027 | | 8.1 | 1.2 | 0.13 | 0.04 | 9.5 | 12 | 9.2 | 1.0 | 19 | 2.8 | 22 |
| **Total Subsystems II-1&3** |||||||||||||
| PH 2009-2015 | | 16 | 0.30 | 0.65 | 0 | 17 | 31 | 19 | NA | 17 | 19 | 35 |
| PH 2015-2021 | | 16 | 0.53 | 0.65 | 0 | 17 | 31 | 20 | NA | 17 | 19 | 35 |
| PH 2021-2027 | | 27 | 5.5 | 0.54 | 0.22 | 33 | 37 | 24 | NA | 34 | 8.6 | 42 |